**Forty-One Types of Physical Quantities in Arbitrary Dimensions**

Venkatraman Gopalan

Department of Materials Science and Engineering, Department of Physics, and Department of Engineering Science and Mechanics, Pennsylvania State University, University Park, PA 16802

It is shown that there are 41 types of multivectors representing physical quantities in non-relativistic physics in arbitrary dimensions within the formalism of Clifford Algebra. The classification is based on the action of three symmetry operations on a general multivector: spatial inversion, $\bar{1}$, time-reversal, $1'$, and a third that is introduced here, namely, wedge reversion, $1^{\dagger}$. It is shown that the traits of "axiality" and "chirality" are not good basis for extending the classification of multivectors into arbitrary dimensions, and that introducing $1^{\dagger}$ would allow for such a classification. Since physical properties are typically expressed as tensors, and all tensors can be expressed as multivectors, this classification also indirectly classifies tensors. Examples of these multivector types from non-relativistic physics are presented.

# 1. Introduction

How many types of (non-relativistic) physical quantities exist in arbitrary dimensions? If the physical quantities are expressed in the formalism of multivectors, an answer provided in this Letter is 41. Physical quantities are widely classified according to the ranks of the tensors representing them, such as scalars (tensors of rank 0), vectors (tensors of rank 1), and tensors of higher ranks [1]. Different tensors transform differently under various spatial and temporal symmetry operations, which provides an additional means to classify them. There is an alternate way to write tensors as *multivectors*, which arise within the formalism of Clifford Algebra (CA) [2,3,4,5]. As simple examples, tensors of ranks 0 and 1 are scalars (S) and vectors(V), which are also components (*blades*) of a general multivector. In CA, one can further continue this sequence and define bivectors(B), trivectors (T), quadvectors (Q) and so on as shown in **Figure 1,** where bivector is a *wedge product* between 2 linearly independent vectors, trivector between 3 such vectors, and so on. A bivector is a *directed* patch of area (i.e. one with a sense of circulation of vectors around its perimeter); a trivector is a directed volume in 3D, a quadvector is a directed hypervolume in 4D and so on. These are examples of *blades*, which are scalars, vectors or wedge product between linearly-independent vectors. For example, the angular momentum, L or magnetic induction, B are truly bivectors of grade 2, though they are conventionally written as axial vectors ***L*** and ***B***, respectively (Bold italics is used for vectors, and plain capitals for other multivectors). Similarly, the torsion of a helix and the phase of a plane wave are trivectors, though they are written normally as scalars. A multivector is an arbitrary sum of such *blades*; for example, M=S+V+B+T+Q is a multivector with five blades of *grade* 0, 1, 2, 3, and 4, respectively. Similarly, the field F=***E***+cB, or the current density J=$(\rho/\varepsilon_o) - c\mu_o$***J***, are multivectors (charge density, $\rho$, $c$ is speed of light in vacuum, permittivity $\varepsilon_o$ and permeability $\mu_o$ of free space). Note that both F and J combine time-even ($E$ and $\rho$) and time-odd (B and $J$) quantities, which is unusual in normal algebra, but perfectly natural in Clifford Algebra. CA allows one to write all four of Maxwell's equations in free space succinctly as one single equation, $(\nabla + [\frac{1}{c}]\partial/\partial t)\text{F} = \text{J}$ in the Newtonian space plus scalar time, *t*, a process called "encoding" that reveals deeper interconnections between diverse laws [2,3]. The real numbers algebra, ordinary vector algebra, complex numbers algebra, quaternions, Lie algebra, are all subalgebras of CA [4,5]. For a reader new to CA, a brief introduction including definitions of multivectors is given in the **Appendix A**.

Hlinka elegantly used group theory to classify these non-relativistic "vectorlike" physical quantities in 3-dimensions (3D) into 8 types [6]. These were time-even (invariant under $1'$) and time-odd (reverses under $1'$) variants of each of the following four types: *neutral*, *polar*, *axial* and *chiral*. Here classical time reversal antisymmetry denoted by $1'$ inverts time, $t \to -t$, and the spatial inversion, denoted by $\bar{1}$ inverts a spatial coordinate $r \to -r$. Neutral and axial type physical quantities are $\bar{1}$-even, while polar and chiral types are $\bar{1}$-odd. In addition, Hlinka imagined these quantities to possess a unique ∞-fold axis in space [6], and considered the 3-dimensional Curie group, $\infty/mm1'$ to represent the quantities as "vectorlike" physical quantities. A mirror $m_{\parallel}$ parallel to this ∞-fold axis was imagined, that would reverse axial and chiral quantities but not the scalar and polar quantities. Thus, the combined actions of $\bar{1}$, $1'$, and $m_{\parallel}$ were used to classify multivectors into the above 8 types. It is shown next that while this classification works in 3-dimensions (3D), it does not translate well into other dimensions. Indeed, physical quantities represented by quadvectors, for example, can only exist in 4- or higher dimensional spaces, and similarly for blades of higher grades. To classify all such multivectors in arbitrary dimensions, we thus need to do two things: first, we need to adopt the framework of Clifford Algebra (CA) within which multivectors arise, and secondly, we have to drop the "axial" and "chiral" traits for classification purposes for reasons described next. While retaining the symmetries of $\bar{1}$ and $1'$ as in Ref. [6] we will need a new symmetry operation that replaces the $m_{\parallel}$ construction. This new antisymmetry will be called wedge reversion, and denoted by $1^{\dagger}$. This classification approach will yield 41 types of multivectors that represent all (non-relativistic) physical quantities in arbitrary dimensions.

First let's note that an axial vector, conventionally defined as the cross product between two polar vectors, $\boldsymbol{V}^{(1)} \times \boldsymbol{V}^{(2)}$ is defined only in 3D (and as an interesting aside, in 7D) [7]. Since an *n*-dimensional vector space cannot contain a blade of grade higher than *n*, one cannot generalize blades of grades other than 3 using axial vectors. Thus, the trait of axiality cannot be generalized to arbitrary dimensions and the concepts of cross-products and axial vectors should therefore be dropped. Secondly, the chirality of a physical quantity depends not only on the grade of the blade, but also the dimension of the ambient space it resides in. To see this, we first note that conventionally, an object is *achiral* if it can be brought into congruence with its mirror image, and *chiral* if it cannot be [8]. If we generalize a mirror in an *n* dimensional space to be an (*n*-1) dimensional hyperplane, then as depicted in **Fig. 2**, in an *n*-dimensional space, only an *n*-dimensional object can be *chiral*. However, the same *n*D object will become *achiral* in a space of dimensionality (*n*+1) or higher. For example, a vector is chiral in 1D but achiral in 2D and higher; a bivector is chiral in 2D and achiral in 3D and higher, and so on. We thus conclude that the trait of chirality (and the $m_{\parallel}$ construction) is also not unique to an object without reference to the dimensionality of the ambient space around the object; hence, chirality too has to be dropped as a trait in uniquely classifying multivectors of arbitrary grade. In essence, a new symmetry operation is needed on par with $\bar{1}$ and $1'$. A good choice turns out to be wedge reversion, $1^{\dagger}$.

## 2. Wedge reversion antisymmetry, $1^{\dagger}$

A brief description of the necessary concepts in Clifford Algebra is first given; a more detailed discussion is given in **Appendix A**. The central concept in CA is the multiplication, or, *geometric product* of two vectors, say **A** and **B**, written as **AB**. "Multiplying" two vectors is possible, for example, if they are expressed in the basis of orthonormal square matrices (such as Pauli and Dirac matrices) as unit vectors. A vector space closed under "geometric product between vectors" is called an *algebra*, and one endowed with a finite vector norm is called the Clifford Algebra or Geometric Algebra [4,5]. For example, one can extend the three-dimensional (3D) vector space spanned by orthonormal basis vectors $\hat{x}, \hat{y}, \hat{z}$ to a $2^3=8$ dimensional CA space spanned by 8 basis vectors, I, $\hat{x}, \hat{y}, \hat{z}, \hat{x}\hat{y}, \hat{y}\hat{z}, \hat{z}\hat{x}, \hat{x}\hat{y}\hat{z}$ (see **Figure 4** in the **Appendix A**). The subspace I of this 8-dimensional CA space is the scalar identity axis that spans all scalars (S), the subspace spanned by the three unit vectors, $\hat{x}, \hat{y}, \hat{z}$ is the vector (V) space, the subspace spanned by the three unit bivectors, $\hat{x}\hat{y}, \hat{y}\hat{z}, \hat{z}\hat{x}$ is the bivector (B) subspace, and the subspace spanned by the unit trivector, $\hat{x}\hat{y}\hat{z}$ is the trivector (T) subspace. A general multivector M in 3D can them be written as a sum of blades in various subspaces; for example, $3+2\hat{x}-5\hat{y}\hat{z}+\hat{x}\hat{y}\hat{z}$ is an arbitrary multivector. Similarly starting from an *n*D vector space, a $2^n$ (or sometimes, less)- dimensional CA space is generated.

The antisymmetry operation of wedge reversion, $1^{\dagger}$ acts on the *wedge product* (∧) between linearly independent vectors (**See Appendix B** for the definition of wedge product). Blades such as bivectors can be written as **A** ∧ **B** between 2 linearly independent vectors, A and B, trivectors as **A** ∧ **B** ∧ **C** between 3 linearly independent vectors, and so on as shown in **Figure 1**. Wedge reversion is not a new operation; it is simply called *reverse*, or *reversion* in Clifford Algebra [2,4]. What is new here is that it is being given the formal status of an antisymmetry, $1^{\dagger}$. The action of wedge reversion, $1^\dagger$, on multivectors is shown in **Figure 1**. Specifically, $1^\dagger(S) = S$, $1^\dagger(V) = V$, and $1^\dagger(V^{(1)} \wedge V^{(2)} \wedge V^{(3)} ... \wedge V^{(n-1)} \wedge V^{(n)}) = V^{(n)} \wedge V^{(n-1)} ... \wedge V^{(3)} \wedge V^{(2)} \wedge V^{(1)}$, where S is a scalar, $V$ is a vector, and $V^{(i)}$ ($i$ is vector index= 1, 2, 3…. n) are $n$ linearly independent vectors. Using orthonormality conditions of the basis vectors stated earlier, one can easily show (**see Appendix C**) that $1^\dagger$ will leave blades of grade $4g$ and $4g+1$ invariant, while reversing the blades of grades $4g+2$ and $4g+3$, where $g$ is a whole number (0, 1, 2, …etc.). Thus, e.g. $1^\dagger$ will leave scalars (grade 0) and vectors (grade 1) invariant, while reversing bivectors (grade 2) and trivectors (grade 3). Since multivectors are sums of blades, the action of $1^\dagger$ on a multivector is clear by noting that it is distributive over addition.

## 3. Group theoretical classification of multivectors

Having unambiguously defined the action of $\bar{1}$, $1'$, and $1^\dagger$ on a blade of any grade, we are now ready to consider the group theoretical aspects of the symmetry group generated by the above three operations, namely, $G = \{1, \bar{1}, 1', 1^\dagger, \bar{1}', \bar{1}^\dagger, 1'^\dagger, \bar{1}'^\dagger\}$, where $g \in G$ is an element of the group $G$. Consider the action of $G$ on any multivector $x \in X$, where $X$ is $2^n$-dimensional CA space. We now consider the *orbit* $O(x) = \{gx \in X : \forall\ g \in G\}$, a set of multivectors obtained by acting on a given multivector $x$ by all elements of the group $G$. Depending on which subset of elements $g \in G$ we pick to create an orbit, one can generate many orbits. These orbits can be uniquely classified based on their *stabilizer subgroups*, $S \leq G$, such that $S(O(x)) = O(x)$. In other words, subgroup $S$ of $G$ consists of all elements of $G$ which leave the orbit invariant. As the group $G$ has 16 subgroups $S$, all orbits of multivectors are classified into 16 classes of orbits as depicted in **Figure 3**. As will be shown below, these 16 classes of orbits form the basis for the classification of the multivectors themselves into 16 categories and within them, 41 types.

**Tables 1** lists the 16 categories of multivectors each represented by a stabilizer subgroup (SS), and within these categories, further categories based on their transformations under $\bar{1}, 1', 1^\dagger$, whether even (*e*), odd (*o*), or mixed (*m*, meaning neither odd nor even). Thus, the SS (Column 2) plus the transformation properties (columns 3, 4, 5) in **Table 1** determine a multivector *type*. These additional *types* were identified by the inspection of the stabilizer subgroup, and identifying the missing symmetry and its possible transformations. For example, the stabilizer subgroup (SS) **1′1**$^\dagger$ must describe multivector blades that are invariant (even, *e*) to both 1′ and 1$^\dagger$. That leaves us with three options for the transformation of the multivectors under the missing antisymmetry, $\bar{1}$, namely, *e, o* or *m*. Of these, the option *e* already corresponds to the multivector type S' with SS of $\bar{\mathbf{1}}$ **1′1**$^\dagger$. That leaves only two unique options for the (SS) **1′1**$^\dagger$ *types*, namely, V' and S'V'. In a similar manner of inspection, all the other types were determined.

In all 41 different types of multivectors are listed, each given a unique letters-based *label* in Column 3. Of these 41, 8 of them are principal types, namely, S, V, B, T, S', V', B', and T'. The action of the three symmetry operations on these eight multivectors is either even or odd, but never mixed. Adopting the terminology of *centric* ($\bar{1}$-even) versus *acentric* ($\bar{1}$-odd), and *acirculant* ($1^\dagger$-even) versus *circulant* ($1^\dagger$-odd) multivectors, we can identify both S and S' in Table 1 to be *centric-acirculant*, both V and V' to be *acentric-acirculant*, both B and B' to be *centric-circulant*, and both T and T' to be *acentric-circulant*. The other 33 multivectors are composed of unique sums of these eight principal multivectors; the action of at least one of the three symmetry operations on these 33 multivectors is mixed. Care is needed in comparing the 8 principal multivector types in this work with those by Hlinka [9]. The *neutral* (types L and N) and *polar* (types T and P) vectors in Ref. [9] do correspond to *centric-acirculant* (types S and S') and *acentric-acirculant* (types V and V') in this work, respectively. However, the axial (M and G) vectors in [9] do *not* correspond to the *centric-circulant* multivectors (B and B'), since they are different grade objects; in this work, axiality as a trait is avoided, and an axial vector is treated no differently from a grade 1 vector of types V or V'. Similarly, chiral pseudoscalars (F and C) in Ref. [9] and *acentric-circulant* multivectors, (T and T') in **Table 1** are different grade objects; the latter make no reference to chirality. In 3D, acentric-circulant multivectors are chiral.

**4. Bidirectors in arbitrary dimensions**

We also make a note of *bidirector*-like quantities in **Table 1**. Hlinka [9] defines a *bidirector* as two opposite vectors $X$ and $-X$ arranged on a common axis at some nonzero distance, $2r$. The bidirector is then represented by the term, $X(r)$-$X(-r)$, where $r$ and $(-r)$ are respectively, the displacement vectors of the vectors $X$ and $-X$, from an origin centered between the two vectors. Note that there is no restriction in picking the directions of the vectors $X$ and $r$; they are independent. Hlinka defines *neutral* (types L and N), and chiral pseudoscalars (F and C) types as bidirectors. The characteristic of these bidirectors is that the vectors composing them are spatially separated and point in opposite ways along a certain direction defined by them; this is unlike a conventional single vector which does not have a well-defined spatial location. **Table 1** lists two types of time-even bidirectors: The S' types with the general form of $(V'(r)$-$V'(-r))$, and the T' type with the general form of $(B'(r)$-$B'(-r))$. There are two types of time-odd bidirectors as well: The S types with the general form of $(V(r)$-$V(-r))$, and the T type with the general form of $(B(r)$-$B(-r))$.

We have to thus generalize Hlinka's definition of bidirectors in arbitrary dimensions as follows: bidirectors are two opposite multivectors, M and -M, arranged on a common axis at some nonzero distance, $2r$. The bidirector is then represented by the term, $M(r) - M(-r)$, where $r$ and $(-$ *r*) are respectively, the displacement vectors of M and -M from an origin centered between the two multivectors.

We can combine bidirectors as well! For example, if $S_{a1}= V_a(r_1)-V_a(-r_1)$ and $S_{b2}= V_b(r_2)-V_b(-r_2)$, (where the each of the subscripts, a, b, 1 and 2, completely define the corresponding vector, and can all be generally linearly independent), then one can naturally define new quantities such as say $S_{a1} \pm S_{b2}$, which now is composed of two bidirectors with a common origin. One could in principle also define such combinations with different origins for different bivectors, such as $S_{a1}(R_{a1}) \pm S_{b2}(R_{b2}))$, where $R_{a1}$ and $R_{b2}$ are the respective origins of bidirectors, $S_{a1}$ and $S_{a2}$. A generalization of the bidirector additions of Type S would thus be, $\sum_k S_k (R_k)$, where $S_k$ is a bidirector of Type S indexed by *k*, and $R_k$ is the location of its origin. This would then give rise to a vector field with specific geometric and symmetry characteristics. One could similarly compose bivector fields, trivector fields, and so on. In general, one could construct multivector fields, $\sum_k Z_k(R_k)$, where $Z_k$ is a bidirector indexed by *k*, composed of an arbitrary multivector, $M_k$ given by $Z_k=M_k(r_k) - M_k(-r_k)$, and centered at $R_k$.

## 5. Examples of different types of multivectors
### 5.1. Helical motion

Examples of non-relativistic multivector types are listed in **Table 1**. Consider a cylindrical helix [10] of radius $\rho$ (note: no relation to charge density defined earlier) and a pitch $2\pi c$ along the helical axis, parametrized by the azimuthal angle $\lambda$ in the plane perpendicular to the helical axis as follows: $q(\lambda) = \rho cos\lambda\, \hat{\gamma}_1 + \rho sin\lambda\, \hat{\gamma}_2 + c\lambda\, \hat{\gamma}_3$, where $\hat{\gamma}_i$, i=1-3, are the orthonormal basis vectors; $q(\lambda)$ is thus a grade-1 homogeneous multivector of Type V'. The *arc length*, $s = |\lambda|\sqrt{\rho^2 + c^2}$ along the helix is a scalar of the Type S'. The tangent vector, $v = q'/|q'|$, as well as the normal vector $\boldsymbol{p} = \boldsymbol{v}'/|\boldsymbol{v}'|$ are vectors of the Type V', where $\boldsymbol{q}' = d\boldsymbol{q}/d\lambda$, and $\boldsymbol{v}' = d\boldsymbol{v}/d\lambda$. The curvature of the path, $K = |\boldsymbol{q}'' \times \boldsymbol{q}'|/|\boldsymbol{q}'|^3 = \rho/(\rho^2 + c^2)$ is a scalar of the type S', where $\boldsymbol{q}'' = d^2\boldsymbol{q}/d\lambda^2$. The *osculating bivector*, $B = (\boldsymbol{v} \wedge \boldsymbol{v}')/|\boldsymbol{v}'|$, is of the type B'. The *torsion*, $\mathcal{T} = (\boldsymbol{v} \wedge \boldsymbol{v}' \wedge \boldsymbol{v}'')/(|\boldsymbol{q}'||\boldsymbol{v}'|^2) = c/(\rho^2 + c^2)$ of the helix is a trivector of the type T'.

Now consider a variant of this helix problem, namely the motion of an object along a cylindrical helical path as a function of time, $t$. If we replace the time-independent variable $\lambda$ in the cylindrical helix example above by $\lambda = \omega t$, where $\omega$ is the angular frequency of the particle moving along this helix, then the action of $(\bar{1}, 1', 1^\dagger)$ on $\boldsymbol{q}(\omega t) = \rho cos\omega t\, \hat{\gamma}_1 + \rho sin\omega t\, \hat{\gamma}_2 + c\omega t\, \hat{\gamma}_3$ is ($o$, $m$, $e$) in **Table 1**, which corresponds to a multivector of the type V'V. The arc length, $s$ and curvature $K$, are still of the type S', while the tangent vector $v$ and the normal vector $p$, are of the type V'V. The osculating bivector B is of the type B'B, while the torsion $\mathcal{T}$ is of the type T'T.

**5.2. Electromagnetism**

Next, examples of the types of multivectors in formulating electromagnetism in CA are presented [3]. In the (3+1)-D formulation, the position vector, $r$, a blade of type V', and the scalar time, $t$, a blade of type S can be combined as a multivector, R= [c]t + r, which is a multivector of Type SV'(S',V). In a similar sense, the spatial vector derivative, $\boldsymbol{\nabla}$, and scalar time derivative, $\partial/\partial t = \partial_t$ can be combined to form a multivector operator $[1/c]\partial_t + \boldsymbol{\nabla}$, which is also of the type SV'(S',V). The terms in the square brackets above and in what follows, are suppressed when expressed in *natural units* for brevity sake, but are understood to be present whenever omitted. The charge density, $\rho$, (type S') and the current density, $\boldsymbol{J}$, (type V) combine to form the multivector electromagnetic source density J=$\rho/[\varepsilon_o] - [c\mu_o]\boldsymbol{J}$, or in natural units, J=$\rho - \boldsymbol{J}$, (type S'V). The electric field **E** (type V') and the magnetic induction bivector B (type B) can be combined into a multivector electromagnetic field F=**E**+[c]B (type V'B(S',T)).

Maxwell's equation in free space condenses to $(\partial_t + \nabla)F = J$, which is a single equation in CA to *encode* all four Maxwell's equations. The left hand side is a geometric product of two multivectors of types SV'(S',V) and V'B(S',T), while the right hand side is a multivector of the type S'V. The blades of different grades collected on each side must equal each other. Expanding this equation by substituting for F and J and solving, we get [3], $\nabla \cdot E + \nabla \wedge B - (\nabla \times B - \partial_t E) + (\nabla \wedge E + \partial_t B) = \rho - J$, where ***B*** is an axial magnetic induction vector while B=$\hat{x}\hat{y}\hat{z}B$ is a magnetic induction bivector. While the right-hand side is a sum of a scalar (type S') and a vector (type V), the left-hand side has a scalar (first term, S'), trivector (2nd term, type T), a vector (3rd term, type V), and a bivector (4th term, type B'). Equating the terms of like multivector grades on the left- and right-hand sides (which should also be of like multivector types), we get the four Maxwell's equations (in natural units), namely, $\nabla \cdot E = \rho$ (Gauss's law), $\nabla \wedge B = 0$ (absence of magnetic monopoles), $(\nabla \times B - \partial_t E) = J$ (Ampere's law with Maxwell's correction), and $(\nabla \wedge E + \partial_t B) = 0$ (Faraday's law). Similarly, the wave equation, $(\nabla^2 - \partial_t^2)F = (\nabla - \partial_t)J$, encodes two Maxwell's wave equations, $(\nabla^2 - \partial_t^2)E = \nabla\rho + \partial_t J$ (multivectors of Type V' on both sides) and $(\nabla^2 - \partial_t^2)B = -\nabla \wedge J$ (multivectors of type B on both sides). In addition, it encodes a third bonus equation, namely, $\partial_t \rho + \nabla \cdot J = 0$, which is a statement of conservation of charge (multivectors of S on both sides). A solution to the encoded wave equation is a plane wave of the type F=F$_o$e$^\Psi$, where $\Psi = \hat{x}\hat{y}\hat{z}(\omega t - \mathbf{k}\cdot\mathbf{r})$ is a trivector of the Type T'T, because $\hat{x}\hat{y}\hat{z}\,\omega t$ is of the type T and $\hat{x}\hat{y}\hat{z}(\mathbf{k}\cdot\mathbf{r})$ is of the Type T'. The field amplitude F$_o$= ***E***$_o$+[c]B$_o$. The fields, F and F$_o$ are of the Type V'B(S',T) as seen before. The generalized electromagnetic energy density, given by $\frac{1}{2}[\varepsilon_o]FF^\dagger = \mathcal{E} + [\frac{1}{c^2}]S$, where $\mathcal{E} = \frac{1}{2}[\varepsilon_o]\mathbf{E}_o^2 + \frac{1}{2}[\mu_o^{-1}]\mathbf{B}_o^2$ is the usual electromagnetic energy density, a scalar of Type S', and $\mathbf{S} = [\mu_o^{-1}](\mathbf{E}_o \times \mathbf{B}_o)$, the Poynting vector of the multivector type V. The corresponding Poynting bivector of type B is S= $[\mu_o^{-1}](\mathbf{E}_o \wedge \mathbf{B}_o)$.

## 6. Conclusions

In conclusion, introducing a new antisymmetry, wedge reversion, $1^\dagger$, in combination with spatial inversion, $\bar{1}$, and classical time-reversal, $1'$, multivectors were classified into 8 principal types, and 41 overall types that classify all physical quantities within the framework of CA. Examples of such multivectors from non-relativistic physics such as helices, helical motion and electromagnetism are presented. Since tensors are widely used to express physical quantities, it is noted that every tensor of rank $r$ in an $n$-D space (therefore, $n^r$ components), can be written as a multivector in a $2^n$ dimensional CA space if $n^r \leq 2^n$ (see **Appendix D**). Thus, the classification of multivectors is equivalent to the classification of tensors. The introduction of two antisymmetries, $1^\dagger$ and $1'$ into conventional crystallographic groups (that already account for $\bar{1}$) form 624 double antisymmetry point groups (DAPG) and 17,803 double antisymmetry space groups (DASG). These have been explicitly listed [11,12]. For a crystal belonging to one of these groups, one can determine the absence, presence, and the form of the 41 multivector types using Neumann's principle [1]. While the development here has focused on non-relativistic physics, we note that that the group theoretic method employed here is blind to the physical meaning of the symmetry operations chosen as long as they generate a group whose elements are all self-inverses and commute with each other.

## APPENDIX *A*: Key concepts in Clifford algebra

We begin with a minimal description of the key concepts in CA essential to following this work; for a more detailed introduction, the reader is referred to Doran and Lasenby [4] and Syngg [5]. The most important concept in CA is that of the multiplication, or, *geometric product* of two vectors, say ***A*** and ***B***, written as ***AB.*** For pedagogical reasons, we begin with three dimensions; extension of the results to *n*-dimensions will then be straightforward. Consider an orthonormal basis set (or a *frame*) of three conventional vectors, $\hat{\gamma}_1, \hat{\gamma}_2, \hat{\gamma}_3$ such that $\hat{\gamma}_1\hat{\gamma}_1 = \hat{\gamma}_2\hat{\gamma}_2 = \hat{\gamma}_3\hat{\gamma}_3 = $ I (normalization condition) and $\hat{\gamma}_i\hat{\gamma}_j + \hat{\gamma}_j\hat{\gamma}_i = 0$ (orthogonality condition, when *i≠j*), where I is identity and the subscripts *i* and *j* each span from 1-3. We can represent these basis vectors as square matrices that satisfy the above relations, e.g. Dirac matrices or Pauli matrices [5], such that a product $\hat{\gamma}_1\hat{\gamma}_2$ for example, simply becomes an elementary matrix multiplication operation of the corresponding matrices for $\hat{\gamma}_1$ and $\hat{\gamma}_2$, which in general is non-commutative, i.e. $\hat{\gamma}_1\hat{\gamma}_2 \neq \hat{\gamma}_2\hat{\gamma}_1$.

With these preliminaries, it is easy to show that arbitrary geometric products of these three basis vectors will result in an expanded algebraic set of $2^3 = 8$ basis vectors as depicted in **Figure 4**: I, $\hat{\gamma}_1, \hat{\gamma}_2, \hat{\gamma}_3, \hat{\gamma}_1\hat{\gamma}_2, \hat{\gamma}_2\hat{\gamma}_3, \hat{\gamma}_3\hat{\gamma}_1, \hat{\gamma}_1\hat{\gamma}_2\hat{\gamma}_3$. We will abbreviate these and group them into subspaces {} as follows: {I}, $\{\hat{\gamma}_1, \hat{\gamma}_2, \hat{\gamma}_3\}$, $\{\hat{\gamma}_1\hat{\gamma}_2, \hat{\gamma}_2\hat{\gamma}_3, \hat{\gamma}_3\hat{\gamma}_1\}$, $\{\hat{\gamma}_1\hat{\gamma}_2\hat{\gamma}_3\}$. We note that this 8-dimensional CA field is composed of the subspace {I} for scalars (also called *grade*-0 *blades*), the vector subspace $\{\hat{\gamma}_i\}$ for the conventional one-dimensional vectors (also called grade-1 blades), the subspace $\{\hat{\gamma}_i\hat{\gamma}_j = \hat{\gamma}_{ij}\}$ for bivectors (grade-2 blades), and subspace $\{\hat{\gamma}_i\hat{\gamma}_j\hat{\gamma}_k = \hat{\gamma}_{ijk}\}$ for trivectors (or grade-3 blades), where *i≠j≠k*. A *multivector* (also called a Clifford number) is an object in this 8-dimensional CA space.

Any product of two vectors will form a *multivector*. For example, if $A = a_i\hat{\gamma}_i$ and $B = b_j\hat{\gamma}_j$ are conventional vectors, then it is straightforward to show that $AB = a_ib_iI + (a_ib_j - a_jb_i)\hat{\gamma}_{ij}$, where $i \neq j$. This is a multivector, $M = AB = \langle M \rangle_0 + \langle M \rangle_2$ with two *blades*, one of grade-0 (denoted $\langle M \rangle_0 = a_ib_iI$) and another of grade-2 (denoted $\langle M \rangle_2 = (a_ib_j - a_jb_i)\hat{\gamma}_{ij}$). A blade is a scalar, a vector, or the wedge product (to be defined shortly) of any number of linearly independent vectors. The *grade* of a blade refers to the number of vectors composing the blade through their wedge product. A general multivector is thus a sum of blades of arbitrary grades; if the grades of all the blades in a multivector are equal, it is called a *homogeneous* multivector.

The coefficient of the first term in M above, $a_ib_i$, can be identified with the conventional *dot* product between the two vectors, $A \cdot B$, and that of the second term with the components of the conventional cross product vector, $A \times B = c_k\hat{\gamma}_k = \epsilon_{kij}\, a_ib_j\hat{\gamma}_k$, where $\epsilon_{kij}$ is the Levi-Civita symbol. Thus one could rewrite as $AB = A \cdot B\, I + (A \times B)_k\hat{\gamma}_{ij}$. One can invert these relationships as, $A \cdot B = (1/2)(AB + BA)$ and (define) $A \wedge B = (1/2)(AB - BA)$ which are the dot product and the *wedge product*, respectively. It is evident from these definitions that $\hat{\gamma}_{12} = \hat{\gamma}_1 \wedge \hat{\gamma}_2$ and $\hat{\gamma}_1\hat{\gamma}_1 = \hat{\gamma}_1 \cdot \hat{\gamma}_1$, and so on for others.

The relationship between the conventional vector cross-product (axial vector) in 3-dimensions and the wedge product is straightforward: $A \times B = -\hat{\gamma}_{123}(A \wedge B)$. Note that in 3D, $A \times B$ is called an *axial* vector that resides in the subspace $\{\hat{\gamma}_i\}$, while $A \wedge B$ is called the bivector that resides in the subspace $\{\hat{\gamma}_{ij}\}$; the above relationship between them defines them as *Hodge Duals* of each other in 3D space (only). Note that axial vectors are in no way special in 3D CA because they reside in the same subspace $\{\hat{\gamma}_i\}$ as the conventional polar vectors. However, axial vectors are typically expressed as wedge products in 3D CA where they live as bivectors in a different subspace $\{\hat{\gamma}_{ij}\}$. While the definition of axial vectors formed through a cross-product between two polar vectors is limited to 3D [13], the *wedge product* between two polar vectors is generalizable to any dimension. Similarly, it can be shown that $\boldsymbol{A} \wedge \boldsymbol{B} \wedge \boldsymbol{C} = \hat{\gamma}_{123} ((\boldsymbol{A}\times\boldsymbol{B}) \cdot \boldsymbol{C})$, where the trivector $\boldsymbol{A} \wedge \boldsymbol{B} \wedge \boldsymbol{C}$ is a Hodge dual of the scalar volume, $(\boldsymbol{A}\times\boldsymbol{B}) \cdot \boldsymbol{C}$ in the 3D space. Thus, in 3D, the Hodge dual of a blade is obtained from its geometric product with the pseudoscalar $\hat{\gamma}_{123}$. These distinctions in 3D are relevant because the antisymmetry $1^\dagger$ reverses the bivector and the trivector in 3D, but *not* the vector (axial or polar) or the scalar.

These ideas can now be generalized to an $n$-dimensional vector space, spanned by orthonormal basis vectors $\hat{\gamma}_i$ ($i \equiv 1, 2, 3, \cdots n$) satisfying the conditions $\hat{\gamma}_i\hat{\gamma}_j + \hat{\gamma}_j\hat{\gamma}_i = 2\delta_{ij}$, where $\delta_{ij}$ is the Kronecker Delta. With the introduction of the geometric product, these $n$-basis vectors expand to a $2^n$ CA space, with subspace $\{I\}$ for *homogeneous multivectors* composed of sums of *grade*-0 blades, subspace $\{\hat{\gamma}_i\}$ for homogeneous multivectors of grade-1, subspace $\{\hat{\gamma}_{ij}\}$ for homogeneous multivectors of grade-2, and so on till $\{\hat{\gamma}_{ijk\ldots(n-1)}\}$ for homogeneous multivectors of grade $(n-1)$ (called *pseudovectors* in dimension $n$), and $\{\hat{\gamma}_{ijk\ldots n}\}$ for homogeneous multivectors of grade $n$, the highest grade blades possible in $n$-dimensions (where $i \neq j \neq k \neq \cdots \neq n$) that is called a *pseudoscalar* in dimension $n$. A blade $\langle M \rangle_p$ of grade $p$ ($\leq n$) in a multivector M in $2^n$ dimensional CA space can always be written as a wedge product $\langle M \rangle_p = V^{(1)} \wedge V^{(2)} \wedge V^{(3)} \wedge \cdots \wedge V^{(p)}$, where $V^{(i)}$ ($i = 1..p$) are $p$ linearly independent grade-1 vectors. The wedge product between any two blades, $\langle M \rangle_p$ and $\langle N \rangle_q$ (grades $p, q \leq n$) in $n$-dimensions can be generalized as $\langle M \rangle_p \wedge \langle N \rangle_q = \langle MN \rangle_{p+q}$. We now have to make a distinction between the *dot* product $\langle M \rangle_p \cdot \langle N \rangle_q = \langle MN \rangle_{|p-q|}$, and the *scalar* product, $\langle M \rangle_p \circ \langle N \rangle_q = \langle MN \rangle_0$; when $p=q$, the dot and scalar products are equal, and when $p \neq q$, the scalar product is zero but the dot product can be non-zero. For any two multivectors, P and Q, each a sum of blades of different grades, the geometric product PQ will contain many blades of different grades. Of these, the sum of blades of the highest grade will be the wedge product P∧Q, and the sum of blades of the lowest grade will be the dot product P·Q. The sum of blades of grade zero will be the scalar product P∘Q. The definition of Hodge dual of a blade can also be generalized to $n$ dimensions by multiplying (i.e. geometric product) of the blade with its pseudoscalar, $\hat{\gamma}_{ijk...n}$, i.e. for example, *P = $\hat{\gamma}_{ijk...n}$P, where * preceding P indicates the Hodge dual of P in the relevant dimension. We finally note that geometric product is distributive over addition.

**APPENDIX *B*.  Wedge product**

The wedge product between $n$ linearly independent vectors is given by the determinant of an $n \times n$ matrix given below:

$$(A \wedge B \wedge C \wedge ...) = \frac{1}{n!} \begin{vmatrix} A & B & C & ... \\ A & B & C & ... \\ A & B & C & ... \\ \vdots & \vdots & \vdots & \ddots \end{vmatrix} \quad (S1)$$

Thus, the dot product and the wedge product between two vectors *A* and *B* can be written down as:

$$(A \wedge B) = \frac{AB - BA}{2} = \frac{1}{2!} \begin{vmatrix} A & B \\ A & B \end{vmatrix}, \quad (S2)$$

and

$$(A \cdot B) = (AB + BA)/2. \quad (S3)$$

From the above definitions, we can deduce that $\hat{x}\hat{x} = \hat{x} \cdot \hat{x}$, and so on for $\hat{y}$ and $\hat{z}$. Similarly, $\hat{x}\hat{y} = \hat{x} \wedge \hat{y}$, and so on for the other basis bivectors, $\hat{y}\hat{z}$ and $\hat{z}\hat{x}$. Finally, $\hat{x}\hat{y}\hat{z} = \hat{x} \wedge \hat{y} \wedge \hat{z}$. Note that the wedge product is non-zero only when the vectors involved are linearly independent.

**APPENDIX *C*.  Wedge reversion**

Wedge reversion $1^\dagger$ is an operation in CA which is generically called *reverse*, or *reversion* or *reversion conjugation*, but is renamed here slightly for uniqueness. (Note the use of *reversion* instead of *reversal*, which we will comment on shortly). The action of $1^\dagger$ on a blade of grade $g$ is simply to reverse the order of the vectors in the wedge product, hence the name given to it. In other words, $1^\dagger \left(\hat{\gamma}_{123\ldots(g-1)g}\right) = \hat{\gamma}_{g(g-1)\ldots321}$. More specifically, given an orthonormal basis, $1^\dagger (\hat{\gamma}_1) = \hat{\gamma}_1, 1^\dagger (\hat{\gamma}_{12}) = \hat{\gamma}_{21} = -\hat{\gamma}_{12}, 1^\dagger (\hat{\gamma}_{123}) = \hat{\gamma}_{321} = -\hat{\gamma}_{123}, 1^\dagger (\hat{\gamma}_{1234}) = \hat{\gamma}_{4321} = \hat{\gamma}_{1234}$, and so on as shown in **Figure 1**.

Given the relation, $(A \wedge B) = \hat{\gamma}_{123} (A \times B)$, and $(A \times B) \cdot C = -\hat{\gamma}_{ijk} (A \wedge B \wedge C)$ in three dimensions, and since $1^\dagger (\hat{\gamma}_{12}) = -\hat{\gamma}_{12}$ and $1^\dagger (\hat{\gamma}_{123}) = -\hat{\gamma}_{123}$, $1^\dagger$ will leave invariant the cross product $(A \times B)$ as well as the scalar $(A \times B) \cdot C$ defined between polar vectors $A$, $B$, and $C$, i.e. $1^\dagger (A) = A$ (and similarly for $B$ and $C$), $1^\dagger (A \times B) = (A \times B)$, $1^\dagger((A \times B) \cdot C) = (A \times B) \cdot C$. We note that the action of that $1^\dagger$ is distributive over addition and multiplication, which is similar to other antisymmetries.

To generalize the action of $1^\dagger$ in $n$-dimensions, lets us denote the pseudoscalar in $n$-dimensional CA as $i_n = \hat{\gamma}_{123\ldots n}$. Then using the orthonormality conditions, one can show that $i_n^2 = (-1)^{n/2}$ (for $n$-even) and $i_n^2 = (-1)^{(n-1)/2}$ (for $n$-odd). The action of $1^\dagger$ on pseudovectors in $n$-dimensions can therefore be derived as $1^\dagger (i_n) = i_n^2 (i_n)$, and $1^\dagger (i_m i_n) = i_n^2 i_m^2 (i_n i_m)$. Using these results, we can show that for a blade $\langle M \rangle_m$ of grade $m$, $1^\dagger \langle M \rangle_m = i_m^2 \langle M \rangle_m$, and $1^\dagger (\langle M \rangle_m \langle M \rangle_n) = i_n^2 i_m^2 (\langle M \rangle_n \langle M \rangle_m)$. Thus $1^\dagger$ will reverse the sign of blades of grades 2, 3, 6, 7, 10, 11$\cdots$ etc., while leaving the blades of grades 0, 1, 4, 5, 8, 9, $\cdots$ etc. invariant. The fact that $1^\dagger$ will reverse the sign of some blades while not reversing that of others, leads us to use the term wedge *reversion*, rather than wedge *reversal*. The former, in particular refers to reversing the order of vectors in a blade, not necessarily the blade itself, as the latter would imply. In particular, the CA of dimensions $n=2$ is isomorphic to the complex algebra where $i_2=\gamma_{12}\equiv\sqrt{-1}$. Noting that $1^\dagger(i_2)=-i_2$, we identify the operation $1^\dagger$ to be isomorphic to complex conjugation in complex algebra. Similarly, the 4-dimensional subspace $\{I, \hat{\gamma}_{12}, \hat{\gamma}_{23}, \hat{\gamma}_{31}\}$ of the $n=3$ CA is isomorphic to the quaternion algebra [14] discovered by Hamilton and whose basis is formed by one real and three imaginary axes. The role of $1^\dagger$ here is again isomorphic to complex conjugation.

**APPENDIX *D*. Tensors expressed as multivectors**

Every tensor of rank $r$ in an $n$-D space (therefore, $n^r$ components), can be written as a multivector in a $2^n$ dimensional space as long as $n^r \leq 2^n$. For example, let $n=4$ and $r=2$. Then, $n^r=2^n=16$, which indicates that a 16-dimensional CA space is in principle sufficient to represent a second rank tensor in 4-dimensional space. If $n^p = 2^n$, then tensors of rank $r<p$ can also be written as multivectors in the $2^n$ dimensional CA space.

As an example, consider the example of a 4×4 second rank tensor, $T=(T_{ij})$, where $i$ and $j$ each range from 0, 1, 2, 3 spanned by an orthonormal basis vectors, $\hat{\gamma}_0, \hat{\gamma}_1, \hat{\gamma}_2, \hat{\gamma}_3$ given by:

$$\hat{\gamma}_0 \equiv \begin{bmatrix} 0 & -\sigma_2 \\ -\sigma_2 & 0 \end{bmatrix}, \quad \hat{\gamma}_1 \equiv \begin{bmatrix} 0 & \sigma_1 \\ \sigma_1 & 0 \end{bmatrix}; \quad \hat{\gamma}_2 \equiv \begin{bmatrix} 0 & \sigma_3 \\ \sigma_3 & 0 \end{bmatrix}; \quad \hat{\gamma}_2 \equiv \begin{bmatrix} I & 0 \\ 0 & -I \end{bmatrix};$$

Where, $\sigma_1 = \begin{pmatrix} 0 & 1 \\ 1 & 0 \end{pmatrix}, \sigma_2 = \begin{pmatrix} 0 & -i \\ i & 0 \end{pmatrix}; \sigma_3 = \begin{pmatrix} 1 & 0 \\ 0 & -1 \end{pmatrix}; I = \begin{pmatrix} 1 & 0 \\ 0 & 1 \end{pmatrix}$. The CA space is spanned by the basis $I, \hat{\gamma}_0, \hat{\gamma}_1, \hat{\gamma}_2, \hat{\gamma}_3, \hat{\gamma}_0\hat{\gamma}_1, \hat{\gamma}_0\hat{\gamma}_2, \hat{\gamma}_0\hat{\gamma}_3, \hat{\gamma}_1\hat{\gamma}_2, \hat{\gamma}_3\hat{\gamma}_1, \hat{\gamma}_2\hat{\gamma}_3, \hat{\gamma}_0\hat{\gamma}_1\hat{\gamma}_2, \hat{\gamma}_0\hat{\gamma}_2\hat{\gamma}_3, \hat{\gamma}_0\hat{\gamma}_1\hat{\gamma}_3, \hat{\gamma}_1\hat{\gamma}_2\hat{\gamma}_3$, and $\hat{\gamma}_0\hat{\gamma}_1\hat{\gamma}_2\hat{\gamma}_3$. The tensor $T$ can then be written as a multivector in this 16-dimensional CA basis as follows:

$$4T = I\,(T_{00} + T_{11} + T_{22} + T_{33}) + i\hat{\gamma}_0(-T_{03} + T_{12} - T_{21} + T_{30}) + \hat{\gamma}_1(T_{03} + T_{12} + T_{21} + T_{30})$$

$$+ \hat{\gamma}_2(T_{02} - T_{13} + T_{20} - T_{31}) + \hat{\gamma}_3(T_{00} + T_{11} - T_{22} - T_{33})$$

$$+ i\hat{\gamma}_0\hat{\gamma}_1(-T_{00} + T_{11} - T_{22} + T_{33}) + i\hat{\gamma}_0\hat{\gamma}_2(T_{01} + T_{10} + T_{23} + T_{32})$$

$$+ i\hat{\gamma}_0\hat{\gamma}_3(T_{03} - T_{12} - T_{21} + T_{30}) + \hat{\gamma}_1\hat{\gamma}_2(T_{01} - T_{10} + T_{23} - T_{32})$$

$$+ \hat{\gamma}_3\hat{\gamma}_1(T_{03} + T_{12} - T_{21} - T_{30}) + \hat{\gamma}_2\hat{\gamma}_3(-T_{02} + T_{13} + T_{20} - T_{31})$$

$$+ i\hat{\gamma}_0\hat{\gamma}_1\hat{\gamma}_2(T_{02} + T_{13} + T_{20} + T_{31}) + i\hat{\gamma}_0\hat{\gamma}_2\hat{\gamma}_3(T_{01} + T_{10} - T_{23} - T_{32})$$

$$+ i\hat{\gamma}_0\hat{\gamma}_1\hat{\gamma}_3(-T_{00} + T_{11} + T_{22} - T_{33}) + \hat{\gamma}_1\hat{\gamma}_2\hat{\gamma}_3(-T_{01} + T_{10} + T_{23} - T_{32})$$

$$+ i\hat{\gamma}_0\hat{\gamma}_1\hat{\gamma}_2\hat{\gamma}_3(T_{02} + T_{13} - T_{20} - T_{31})$$

While the above expression for *T* is obtained from a straight-forward matrix decomposition in the basis of the orthonormal matrices given above, a conceptual transition is required in transitioning from the "matrices" $\gamma_i$ to "vectors" $\hat{\gamma}_i$ in CA. In particular, the rules for linear transformation of matrices versus those for vectors in CA differ, and appropriate caution must be exercised in working out the appropriate correspondences between the two. As a simple example, a rotation of a 2×2 matrix by an angle $\theta$ in the 1-2 plane is performed in conventional tensor algebra by the transformation matrix, $\begin{pmatrix} \cos\theta & \sin\theta \\ -\sin\theta & \cos\theta \end{pmatrix}$, while such a rotation of a corresponding "vector" in CA would require a rotor operator $R = \cos\left(\frac{\theta}{2}\right) - \hat{\gamma}_1\hat{\gamma}_2 \sin\left(\frac{\theta}{2}\right)$.

V.G. gratefully acknowledges support for this work from the National Science Foundation grant number DMR-1807768 and the Penn State NSF-MRSEC Center for Nanoscale Science, grant number DMR 1420620. Brian K. VanLeeuwen first proposed classifying vectors using stabilizer subgroups in 2014. Discussions during that period with Brian K. VanLeeuwen, Daniel B. Litvin, and Mantao Huang on problems with defining a "cross-product reversal" antisymmetry led V.G. to Clifford Algebra and wedge reversion. Discussions with John Collins, Haricharan Padmanabhan, Jason Munro, Ismaila Dabo, Chaoxing Liu, Jeremy Levy and Susan Sinnott are acknowledged.

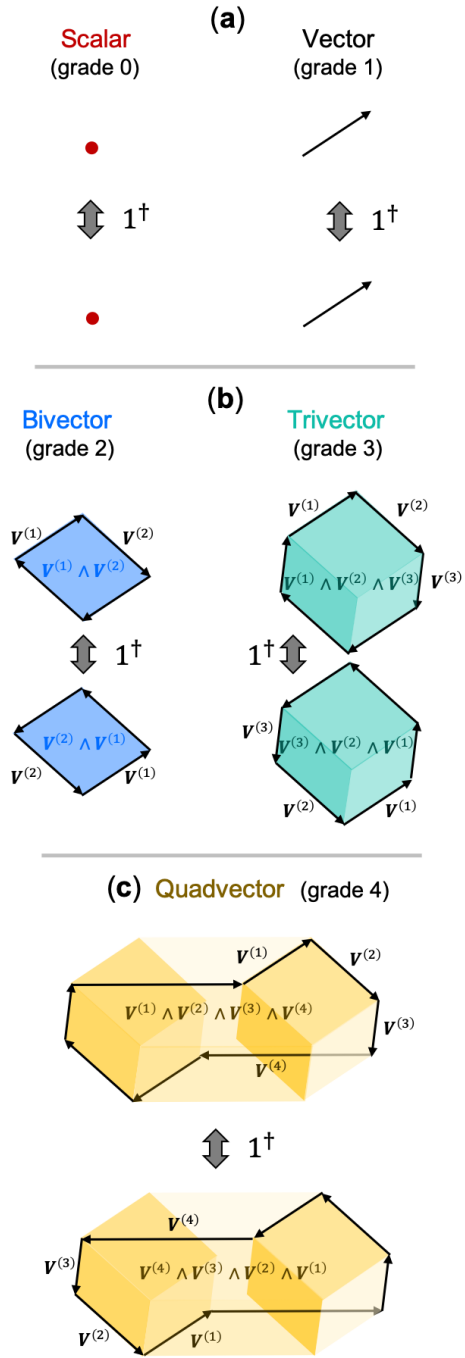

**FIG. 1**. The action of wedge reversion, $1^\dagger$ on (a) A scalar (red dots) and a vector (black arrows), (b) A bivector (the blue patch of area, $V^{(1)} \wedge V^{(2)}$) and a trivector (the sea-green 3D volume, $V^{(1)} \wedge V^{(2)} \wedge V^{(3)}$), and (c) a quadvector (the yellow hypervolume in 4D, $V^{(1)} \wedge V^{(2)} \wedge V^{(3)} \wedge V^{(4)}$). Panel

(c) has to be imagined as a 4D object. Scalars, vectors and wedge products (∧) between linearly independent vectors, $V^{(i)}$ indexed by natural numbers, $i$, are called *blades* and their grades are indicated above. Blades of grades $4g$ and $4g+1$ remain invariant, while those of grades $4g+2$ and $4g+3$ reverse under the action of $1^\dagger$, where $g$ is 0, 1, 2, 3…etc.

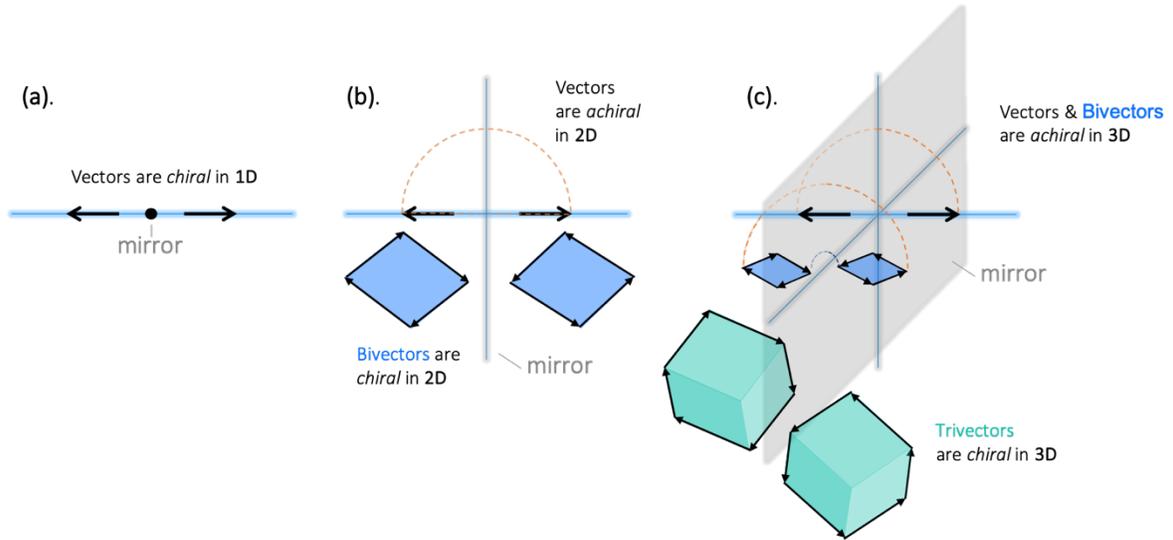

**Fig. 2.** (*a*) A vector and its mirror image (black arrows) cannot be congruently overlapped in 1D, and hence it is chiral. (*b*) However, a vector and its mirror image can be overlapped congruently in 2D when pivoted with the arrow head along the broken light-orange line trajectory, indicating it is achiral. However, a bivector and its mirror image (light blue parallelograms with right-handed and left-handed circulations around their perimeters) cannot be congruently overlapped in 2D indicating they are chiral in 2D. (*c*) A vector and a bivector are both achiral in 3D as indicated by light-orange broken lines indicating the suggested trajectory for overlapping the objects and their mirror images. However, a trivector and its mirror image in light green with vector circulations shown cannot be congruently overlapped in 3D, and hence it is chiral; it will no longer be chiral in 4D and higher dimensions. In general in *n*-dimensions (*n*D), a chiral object can only be *n*D, and it is no longer chiral in (*n*+1)-D or higher, where *n* is a natural number.

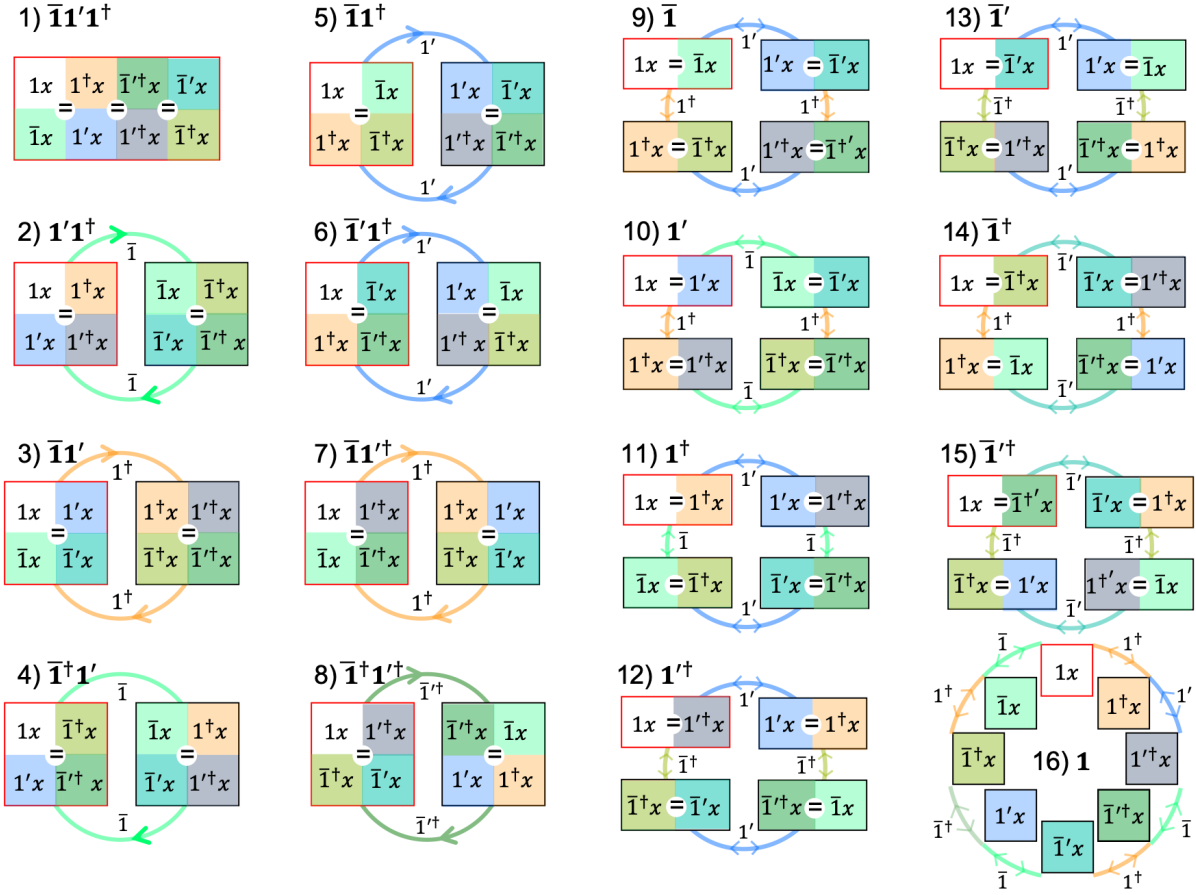

**Fig. 3**: Sixteen orbits of the antisymmetry group $G = \{1, \bar{1}, 1', 1^\dagger, \bar{1}', \bar{1}^\dagger, 1'^\dagger, \bar{1}'^\dagger\}$, representing the action of the elements of the group (each represented by its own color) on a multivector, $x$. The orbits are labeled by the generating elements of their stabilizer subgroups (SS). For example, orbit 5) $\bar{1}\,1^\dagger$ is identified uniquely by its stabilizer subgroup $S=\{1, \bar{1}, 1^\dagger, \bar{1}^\dagger\}$, generated by the generating elements $\bar{1}$ and $1^\dagger$. The squares and rectangles with a red outline represent the actions of the stabilizer subgroups on $x$. The quantities in the squares adjoining any equal-to sign, "=", are equal to each other.

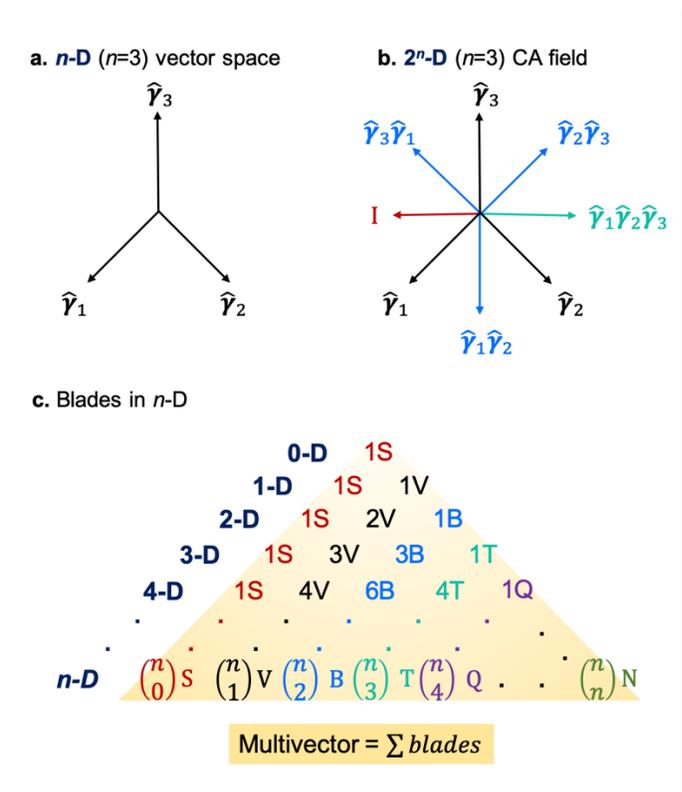

**Fig. 4**: (a) Example of an $n=3$-dimensional vector space with orthonormal basis vectors $\hat{\gamma}_1, \hat{\gamma}_2, \hat{\gamma}_3$, expands under the operation of geometric product (multiplication) of vectors to a $2^n=8$-dimensional algebraic vector space, (b), that is closed under geometric product. Objects in this space are called multivectors (or Clifford numbers), whose algebra is called the Clifford Algebra (CA). A blade is a scalar, a vector, or the wedge product of any number of vectors. The grade of a blade refers to the number of vectors composing the blade. The number of blades of different grades forming the basis for a $2^n$-dimensional CA is given by the Pascal's triangle, (c). The correct sequential notation for blades in increasing order of grade is *scalar* (S), *vector* (V), *bivector* (B), *trivector* (T), *quadvector* (Q),… *blade of grade n* (N). A general multivector in CA is a blade or a sum of blades.

**Table 1**: The classification of 41 types of multivectors (column 1) based on 16 stabilizer subgroups (column 2) of the group, $G = \{1, \bar{1}, 1', 1^\dagger, \bar{1}', \bar{1}^\dagger, 1'^\dagger, \bar{1}'^\dagger\}$. A multivector *type* is defined as a unique combination of entries in columns 2, 4, 5, and 6. The bolded entries (**1**, **2**, **4**, **6**, **8**, **10**, **12**, **14**) in column 1 are the eight principal multivector types. Column 2 lists the generating elements for the stabilizer subgroups (SS) for the 16 orbit types given in Figure 2. A horizontal grey or a white band represents all the rows to which the corresponding SS applies. Suggested notation for different multivector types is introduced in column 3. Columns 4-6 present the action of three symmetry operations, $\bar{1}, 1'$, and $1^\dagger$, on these 41 multivectors as either even (*e*), odd (*o*), or mixed (*m*). Column 7 presents the possible grades of the blades whose sum forms the corresponding multivector type. Colum 8 presents some examples of multivector types. A multivector of the type VT'=V+T' etc. A multivector labeled SB'(S',B) is a mandatory sum of vectors G and L, along with optional additions of types B or S' or both. Conventional vectors are presented in ***bold italics***. Multivector labels are presented as CAPITAL letters without italics or bold. For example, note the distinction between B, a notation for a *general* time-odd bivector (column 3), versus ***B***, an axial magnetic induction vector. Similarly, ***r***, ***r***$_i$ (*i*=1, 2,3): position vector; ***P***, ***P***$_i$: polarization; ***E***: electric field; ***v***: velocity; ***J***: current density; ***p***: momentum; ***H***: magnetic field; ***B***: Magnetic induction; *t*: time. The * as in *****B***=$\hat{x}\hat{y}\hat{z}$***B*** indicates a Hodge dual of ***B*** in 3D, and similarly for others. In 3D, the Hodge dual of a vector is a corresponding bivector and vice versa.

| # | SS | Label | Action of $\bar{1}$ | Action of $1'$ | Action of $1^\dagger$ | Grades | Examples of Multivectors |
|---|---|---|---|---|---|---|---|
| 1 | $\bar{1}1'1^\dagger$ | S' | e | e | e | 4g | $t^2$, **P**(**r**)-**P**(-**r**), V'(**r**)-V'(-**r**) |
| 2 |  | V' | o | e | e | 4g+1 | **r**, **P**, **E**, $P_1 \wedge P_2 \wedge P_3 \wedge P_4 \wedge P_5$, |
| 3 | $1'1^\dagger$ | S'V' | m | e | e | 4g, 4g'+1 | S'+V' |
| 4 |  | B' | e | e | o | 4g+2 | $\nabla \wedge \boldsymbol{E}$, $\boldsymbol{r} \wedge \boldsymbol{P}$, $P_1 \wedge P_2 \wedge P_3 \wedge P_4 \wedge P_5 \wedge P_6$ |
| 5 | $\bar{1}\,1'$ | S'B' | e | e | m | 4g, 4g'+2 | S'+B' |
| 6 |  | T' | o | e | o | 4g+3 | $r_1 \wedge r_2 \wedge r_3$, $\boldsymbol{r} \wedge B'$, B'(**r**)-B'(-**r**) |
| 7 | $\bar{1}^\dagger 1'$ | S'T' | m | e | m | 4g, 4g'+3 | S'+T' |
| 8 |  | S | e | o | e | 4g | $t$, V(**r**)-V(-**r**), $r_1 \wedge r_2 \wedge r_3 \wedge \boldsymbol{p}$ |
| 9 | $\bar{1}1^\dagger$ | S'S | e | m | e | 4g, 4g' | S'+S |
| 10 |  | V | o | o | e | 4g+1 | **v**, **J**, **p**, $r_1 \wedge r_2 \wedge r_3 \wedge r_4 \wedge \boldsymbol{p}$ |
| 11 | $\bar{1}'1^\dagger$ | S'V | m | m | e | 4g, 4g'+1 | S'+V |
| 12 |  | B | e | o | o | 4g+2 | M=*M, *B, L=$\boldsymbol{r} \wedge \boldsymbol{p}$, $r_1 \wedge r_2 \wedge r_3 \wedge r_4 \wedge r_5 \wedge \boldsymbol{p}$ |
| 13 | $\bar{1}1'^\dagger$ | S'B | e | m | m | 4g, 4g'+2 | S'+B |
| 14 |  | T | o | o | o | 4g+3 | $r_1 \wedge r_2 \wedge \boldsymbol{p}$, $\boldsymbol{r} \wedge B$, B(**r**)-B(-**r**) |
| 15 | $\bar{1}^\dagger 1'^\dagger$ | S'T | m | m | m | 4g, 4g'+3 | S'+T |
| 16 |  | SB'(S',B) | e | m | m | 4g, 4g'+2 | B'+S, S+B'+B, S+S'+B', S+S'+B+B' |
| 17 | $\bar{1}$ | SB | e | o | m | 4g, 4g'+2 | S+B |
| 18 |  | B'B | e | m | o | 4g+2, 4g'+2 | B'+B |
| 19 |  | V'B(S',T') | m | e | m | all | B'+V', V'+B'+T, S'+V'+B', S'+V'+B'+T' |
| 20 | $1'$ | V'T' | o | e | m | 4g+1, 4g'+3 | V'+T', V'+T' |
| 21 |  | B'T' | m | e | o | 4g+2, 4g'+3 | B'+T' |
| 22 |  | SV'(S',V) | m | m | e | 4g, 4g'+1 | S+V', S+V+V', S+S'+V', S+S'+V+V' |
| 23 | $1^\dagger$ | V'V | o | m | e | 4g+1, 4g'+3 | V'+V |
| 24 |  | SV | m | o | e | 4g, 4g'+1 | S+V |
| 25 |  | V'B(S',T) | m | m | m | all | V'+B, V'+B+T, S'+V'+B, S'+V'+B+T |
| 26 | $1'^\dagger$ | BT | m | o | o | 4g+2, 4g'+3 | B+T |
| 27 |  | V'T | o | m | m | 4g+1, 4g'+3 | V'+T |
| 28 |  | VB'(S',T) | m | m | m | all | V+B', V+B'+T, V+B'+N, S'+V+B'+T |
| 29 | $\bar{1}'$ | VT | o | o | m | 4g+1, 4g'+3 | V+T |
| 30 |  | B'T | m | m | o | 4g+2, 4g'+3 | T+B' |
| 31 |  | ST'(S',T) | m | m | m | 4g, 4g'+3 | S+T', S+T+T', S+S'+T', S+S'+T+T' |
| 32 | $\bar{1}^\dagger$ | T'T | o | m | o | 4g+3, 4g'+3 | T+T' |
| 33 |  | ST | m | o | m | 4g, 4g'+3 | S+T |
| 34 |  | VB | m | o | m | 4g+1, 4g'+2 | V+B |
| 35 | $\bar{1}'^\dagger$ | BT' | m | m | o | 4g+2, 4g'+3 | B+T' |
| 36 |  | VT' | o | m | m | 4g+1, 4g'+3 | V+T' |
| 37 |  | S'VBT' | m | m | m | varied | V+B+T', S'+V+B, S'+V+B+T' |
| 38 |  | W | m | m | m | varied | e.g. S'VBT' + any from #16 to #33 |
| 39 |  | X | m | m | o | varied | e.g. BT'+ TT'+B'T+BT+B'T'+BB'+T+T'+B+B' |
| 40 | 1 | Y | m | o | m | varied | e.g. VB+ST+VT+BT+SV+SB+S |
| 41 |  | Z | o | m | m | varied | e.g. VT'+TT'+VT+V'T+VV'+V'T'+V+V'+T+T' |

# References

vxg8@psu.edu

Press, Natick, MA, 2003).